\documentclass[12pt]{article}
\usepackage{epsfig,amsfonts,amssymb}
\usepackage[utf8]{inputenc}
\usepackage{hyperref}

\usepackage{comment}
\usepackage{cite}
\input epsf.sty
\usepackage{cite}

\usepackage{authblk}
\usepackage{setspace}
\usepackage{amsmath,amssymb,amsfonts,amsthm,mathrsfs}
\usepackage{graphicx}
\usepackage{enumerate} 
\usepackage{comment}

\newcommand{\bigO}{\mathcal{O}}

\def\ZZZ{{\hbox{ Z\kern-1.6mm Z}}}
\def\RRR{{\hbox{ R\kern-2.4mm R}}}
\def\CCC{{\hbox{ C\kern-2.0mm C}}}
\def\zzz{{\hbox{z\kern-1mm z}}}

\newcommand{\qeq}{{\hbox{=\kern-2.3mm ? \kern.5mm }}}
\renewcommand{\qeq}{=}

\newcommand{\s}{\mathcal{S}}

\newcommand{\be}{\begin{equation}}
\newcommand{\ee}{\end{equation}}
\newcommand{\ben}{\begin{eqnarray}\displaystyle}
\newcommand{\een}{\end{eqnarray}}

\def\one{{\hbox{ 1\kern-.8mm l}}}
\def\zero{{\hbox{ 0\kern-1.5mm 0}}}

\newcommand{\bea}[1]{\begin{eqnarray}\label{#1} }
\newcommand{\eea}{\end{eqnarray}}

\usepackage{bm}
\usepackage[table]{xcolor}

\begin{document}

\baselineskip 24pt

\begin{center}

{\Large \bf Supertranslations in Higher Dimensions Revisited}


\end{center}

\vskip .6cm
\medskip

\vspace*{4.0ex}

\baselineskip=18pt

\centerline{\large \rm Ankit Aggarwal}

\vspace*{4.0ex}

\centerline{\large \it The Institute of Mathematical Sciences, HBNI}
\centerline{\large \it   CIT Campus, Chennai-600 113, India.}


\vspace*{1.0ex}
\centerline{\small E-mail:  aankita@imsc.res.in}

\vspace*{5.0ex}

\centerline{\bf Abstract} \bigskip

In this paper, we revisit the question of identifying Soft Graviton theorem in higher (even) dimensions with Ward identities associated with Asymptotic symmetries. Building on the prior work of \cite{strominger}, we compute, from first principles, the  (asymptotic) charges associated to Supertranslation symmetry in higher even dimensions and show that (i) these charges are non-trivial, finite and (ii) the corresponding Ward identities are indeed the soft graviton theorems.

\vfill \eject

\baselineskip=18pt

\section{Introduction}
Asymptotically flat spacetimes are those which approach flat spacetimes `far away' from the matter sources. A precise mathematical definition of asymptotic flatness involves specifying the boundary conditions- the rate at which the metric of the spacetime approaches the flat metric near its boundary. There is no unique prescription to specify these boundary conditions. Any particular choice is based on the following two physically motivated guiding principles (see the introduction of  \cite{Wald}). Firstly, the conditions should be weak enough to allow for physically interesting solutions like black holes and gravitational radiation. Secondly, the conditions should be strong enough to ensure that the physically interesting notions like total mass or total radiated energy are finite and well defined. \\

The group of non-trivial transformations which leave the form of the specified boundary conditions invariant is known as the asymptotic symmetry group (ASG).
The work of Bondi, Metzner, and Sachs (\cite{Bondi, Sachs1, Sachs2})  in the 1960s showed that the ASG of asymptotically flat spacetimes in four dimensions is the infinite dimensional BMS group. It is a  semidirect product of supertranslations (angle dependent translations along null infinity), and the Lorentz group. Curiously, however, the studies of asymptotically flat spacetimes in higher dimensions  (\cite{hollands1, hollands2, Wald, tanabe}) have concluded that the supertranslations do not form a part of the ASG in $d>4$. This negative result is a consequence of using stringent boundary conditions that were guided by the physical considerations of the kind mentioned above. A remarkably precise description of these considerations was provided in \cite{Wald} for even $d>4$. 
 Firstly, it was shown that supertranslations are related to the memory effect and that in higher even dimensions, memory effect is not present 
(to the order at which the tidal effects due gravitational radiation are seen). 
 This was supplemented by the statement (see conclusions of \cite{Wald}) that if weaker boundary conditions are imposed to allow for supertranslations, it would lead to an ill-defined charge (flux) for supertranslations (due to the divergence of symplectic current). Both of these arguments do not hold in four dimensions. In fact, supertranslations are inevitable in $d=4$. Boundary conditions disallowing supertranslations automatically disallow all the generic radiative solutions (see e.g. \cite{Wald}).  \\
 
 However, the absence of BMS (supertranslations) presents a puzzle. In a remarkable program initiated by Strominger \textit{et al} (\cite{stromingeryang, strominger1, stromingerqed1, stromingerqed2, stromingerqedsub,   strominger2, stromingergravitysub, laddha, campi1, laddhaqed}; see \cite{stromingerlectures} for a review)
 , soft theorems have been understood to be the Ward identities associated with the asymptotic symmetries in gauge and gravitational theories.  
  The Ward identities of BMS (and suitable extensions thereof) have been shown to be equivalent to the leading (subleading) soft graviton theorem in four dimensions (\cite{strominger2, stromingergravitysub, laddha, campi1}). 
   In particular, in \cite{strominger2}, a  diagonal subgroup of the direct product of BMS group at future ($\mathcal{I}^+$) and past null infinity ($\mathcal{I}^-$) was identified as a symmetry group of quantum $\s$-matrix. The Ward identity for this symmetry group was then shown to be the leading/Weinberg's soft graviton theorem \cite{weinberg} in four dimensions. \\

 Furthermore, it is well understood that the soft graviton theorems exist in all dimensions \cite{senall1, senall2, senall3}. Thus, a natural question to ask is whether or not the soft graviton theorems in higher dimensions are also statements about the symmetries of the quantum gravity $\s$-matrix.  
   Motivated by this, the authors in \cite{strominger} argued for the existence of supertranslations in even $d=2m+2>4$.  
 Starting from the leading soft graviton theorem, they wrote it as a Ward identity for the $\s$-matrix and read off a (conserved) charge. By proposing suitable commutation relations for radiative degrees of freedom, this charge was shown to generate supertranslations in even $d>4$ ($m>1$). Hence, results of \cite{strominger} present a very compelling argument in favor of supertranslations in higher even dimensions. However, to unambiguously establish the existence of supertranslations in higher dimensions, one needs to show that: (i) there exist suitable boundary conditions allowing them; (ii) the associated charge is finite and well-defined and (iii) the corresponding Ward identity is the leading soft graviton theorem. This wasn't established in \cite{strominger} 
  and the works of \cite{hollands1, hollands2, Wald, tanabe} seem to be at odds with this expectation. Relaxing the boundary conditions would allow supertranslations but would lead to an ill-defined charge as was pointed out in \cite{Wald} (see paragraph 2 above).    \\ 
 
 \emph{Thus, there is a tension between the results of  classical GR 
  and those of \cite{strominger}.}
 \\

 In this paper, we resolve this conundrum by proposing some additional boundary conditions near the boundaries of null infinity. As we shall see, these boundary conditions will lead to a finite, well-defined charge for supertranslations in higher even dimensions. We will also see that the boundary conditions that we propose can be thought of as analogs of CK (Christodoulou Klainermann)-constraints in four dimensions (\cite{strominger1}). Doing this helps in getting the right count of the number of independent leading soft graviton theorems (one). We will work in linearized gravity throughout the paper.
 \\

 The outline of the paper is as follows. In section 2, we give the preliminaries needed for our work and explain the notation. Then, we do a detailed analysis of the six-dimensional case in section 3 where we compute the charges for supertranslations and give the generalized CK constraints in six dimensions.  
Our results are generalized to arbitrary higher even dimensions in section 4. In section 5, we discuss the potential problems in going from linear to the non-linear regime. We conclude with a summary and future outlook in section 6.

\section{Preliminaries} \label{sec2}
Our attention will be limited to arbitrary even dimensions because the notion of conformal null infinity breaks down in odd dimensions (\cite{waldodd}). We will mostly use the notation of \cite{strominger} and work in linearized gravity coupled to massless matter in $d=2m+2$ dimensions. Linearization simplifies the analysis and is justified when there are no `hard' gravitons (gravitons with non-zero momentum) present in the external states. 
  This will, in particular, mean that we will only keep the terms linear in metric fluctuations in the  charge (the intermediate computations like those of symplectic potential would require us to keep terms quadratic in metric fluctuations since they contribute linearly to the charge). When there is no scope of confusion, the contraction of indices is denoted with a dot.
We are interested in spacetimes that are asymptotically flat at both future and past null infinity.  For concreteness we focus on future null infinity; similar considerations apply to the past null infinity. Most of the content of this section (except section 2.4) can be found in \cite{strominger} with more details. We work in units where $8\pi G=1$ which is different from \cite{strominger} where $32\pi G=\kappa^2$.
\subsection{Metric and Bondi Gauge} \label{metric}

We work in the Bondi coordinates $(u,r,z^A)$, where $u$ is the retarded time, $r$ is the radial coordinate and $z^A$ are the coordinates on the sphere. The linearized $d$-metric is parameterized as  
\be
ds^2 = M  du^2 - 2  du dr + g_{AB} dz^A dz^B -2 U_A dz^A du , \label{4metric}
\ee
 
where $A, B=1,...2m$ denote the sphere indices.   
The inverse metric is given by:
\be
g^{\mu\nu}= \begin{pmatrix}
0 & -1 & 0\\
-1& -M &-U^B\\
0 & -U^A & g^{AB} 
\end{pmatrix} 
\ee

We will assume  $M$, $U_A$, and $g_{AB}$ admit an expansion near $\mathcal{I}^+$ of the form:
\begin{gather} \label{eq:e1} \notag
M=-1+\sum_{n=1}^{\infty}\frac{M^{(n)}(u,z)}{r^n},  \hspace{.5 in} U_A=\sum\limits_{n=0}^\infty \frac{U_A^{(n)}(u,z)}{r^n},
\\  g_{AB}=r^2\gamma_{AB}+\sum_{n=-1}^{\infty}\frac{C_{AB}^{(n)}(u,z)}{r^n}.   
\end{gather}
There is an additional determinant condition in the Bondi gauge 
 \be
\det(g_{AB})= r^{2m}\det(\gamma_{AB})
\ee
 where $\gamma_{AB}$ is the round metric on $S^{2m}$.
This condition, in linearized theory implies that all the $C_{AB}^{(n)}$ are traceless
\be
\gamma^{AB}C_{AB}^{(n)}=0.
\ee

 \subsection{Boundary Conditions and Constraints From Einstein's Equations} \label{BC}
 Following \cite{strominger},  the 
 boundary conditions  for asymptotically flat spacetimes in arbitrary even dimensions, are taken to be
\be \label{bc1}
g_{uu}=-1+\bigO(r^{-1}), \hspace{.3 in} g_{ur}=-1+\bigO(r^{-2}), \hspace{.3 in} g_{uA}=\bigO(1), \hspace{.3 in} g_{AB}=r^2\gamma_{AB}+\bigO(r).   
\ee
We also require
\be \label{bcric} 
R_{uu}=\bigO(r^{-2m}), \hspace{.5 in} R_{ur}=\bigO(r^{-2m-1}), \hspace{.5 in} R_{uA}=\bigO(r^{-2m}),  
\ee
\be \label{bc2}
R_{rr}=\bigO(r^{-2m-2}), \hspace{.5 in} R_{rA}=\bigO(r^{-2m-1}), \hspace{.5 in}   R_{AB} = \bigO(r^{-2m}).   
\ee

Linearized Einstein's equations 
\be \label{Ricci}
R_{mn} = \bar{\Box} h_{mn} - 2\bar{\nabla}_{(m} \bar{\nabla}\cdot h_{n)} = T_{mn}  .
\ee
 determine metric components in terms of the free radiative as well as matter data. In Bondi coordinates, we get the following constraints:
 
The boundary condition on  $R_{ur}$ reads
\be
 -\frac{ n(n+1-2m)}{2}M^{(n)}  +\frac{  (n-1)}{2} D^A U_A^{(n-1)}=0, \hspace{.5 in}  0 \leq n \leq 2m-2.   
\ee
The boundary condition on $R_{rA}$ reads
\be
 \frac{(n+2)(n+1-2m)}{2}U_A^{(n)} - \frac{(n+1)}{2}D^BC_{BA}^{(n-1)}=0, \hspace{.5 in} 0\leq n \leq 2m-2.   
\ee
The $R_{AB}$ equations lead to
\be \label{eeC}
\partial_uC_{AB}^{(-1)}=0. 
\ee
This should be contrasted with the four dimensions, where (\ref{eeC}) isn't valid and $C_{AB}^{(-1)}$ is the free radiative data which, generically, can depend on $u$.\\
$R_{uu}$ equations give
\be \label{constraint}
\frac12 [D^2- 2(m-1)] M^{(2m-2)}  + \partial_u D^A U_A^{(2m-2)} + m  \partial_u M^{(2m-1)}  +  T^{M(2m)}_{uu} =0.   
\ee

where $T_{uu}^{M(2m)}$ is the $\mathcal{O}(r^{-4})$ component of matter stress-energy tensor which is assumed to have the same fall off behaviour as Ricci tensor.
$C_{AB}^{(m-2)}$ is free radiative data (see \cite{strominger}) as can be seen from saddle point approximation.

\subsection{Supertranslations in higher even dimensions}
In four dimensions, supertranslations form an infinite dimensional group (which is an abelian subgroup of the 
 the BMS group). They are characterized by an arbitrary function on the sphere. In higher even dimensions, for the boundary conditons under consideration (see section \ref{BC}), there is an analogus infinite dimensional group characterized an arbitrary function of the sphere coordinates, $f(z)$ \footnote{It is also an abelian subgroup of the analogue of BMS in higher even dimensions. }(see \cite{strominger}). This will be referred to as supertranslations.  Near~$\mathcal{I}^+$, supertranslations are generated by the vector field 
\be 
\xi =f\partial_u -\frac{1}{r}\gamma^{AB}D_A f \partial_B +\frac{1}{2m} D^2 f \partial_r+\dots   
\ee
where $\dots$ refer to subleading terms in $r$. It can be easily checked that the above vector field preserves the form of the metric near $\mathcal{I}^+$, making it a valid asymptotic symmetry.\\
 The effect of the supertranslations is to shift $C_{AB}^{(-1)}$ according to 
\be\label{st}
\delta_{\xi} C_{AB}^{(-1)}
=\frac{1}{m}D^2f\gamma_{AB} -(D_AD_B+D_BD_A)f. 
\ee
\\

For the remaining $C_{AB}^{(n)}$s, 

\be
\delta_{\xi} C_{AB}^{(n \geq 0)}=\mathcal{O}(C).
\ee

 Hence, in the linearized theory \footnote{Since we are working in the linearized theory, we have ignored the terms homogeneous in metric perturbations because these will lead to non-linear terms in charge as will become clear in the  following sections.}

\be
\delta_{\xi} C_{AB}^{(n \geq 0)}=0. 
\ee

It is important to note that in four dimensions the free radiative data is $C^{(-1)}_{AB}$ which is affected by the supertranslations within the linearized regime. This is not so for higher dimensions where the free radiative data is $C^{(m-2)}_{AB}$ (as noted above) which remains unaffected by supertranslations in the linearized regime. 

\subsection{Covariant Phase Space Formalism} \label{covariant}

We will be using covariant phase space formalism to compute the charge. This subsection enlists the steps involved in  this computation (see \cite{ashtekar, lee, zoupas, compere } for a detailed discussion). The first step is to  evaluate symplectic potential
\be
\Theta_{t}(\delta) := \int_{\Sigma_t} dS_a \theta^a,
\ee

where
\be
\theta^a :=\frac{1}{2} \sqrt{g} \big( g^{bc} \delta \Gamma^a_{bc} - g^{a b} \delta \Gamma^c_{c b} \big)
\ee
and $\Sigma_t$ is a $t=constant$ slice.
Since $t:=r+u$, the component of interest will be $\theta^t = \theta^r+\theta^u$.  The limit $t \to \infty$ with $u$ constant takes us to $\mathcal{I}^+$ which is the surface on which the charge (flux) is computed.  The variable $r$ will be understood as given by $r=t-u$ while performing the integration on the spatial slice. \\

Using symplectic potential, the standard covariant phase space symplectic form can be computed

\be
\Omega_{t,g}(\delta ,\delta'  ) := \int_{\Sigma_t} dS_a \omega_g^a(\delta,\delta'), \label{Ot}
\ee

where 
\be
\omega_g^a(\delta,\delta')=\delta \theta^a(\delta')-\delta' \theta^a(\delta)
\ee
The charge is given by integrating

\begin{equation}
\Omega(\delta_{\xi},\delta)\ =:\ \delta Q_{\xi}.  \label{defHV}
\end{equation}

\section{Linearized gravity in six dimensions} \label{sec3}

Before considering arbitrary even dimensions, it will be useful to first study the case of six dimensions in detail which is done in this section. In section 3.1, we state the fall-offs and the constraints imposed by the Einstein's equations. In section 3.2, symplectic potential is evaluated which turns out to be divergent as $t\rightarrow\infty$.
We will then consider the problematic (diverging as t$\rightarrow\infty$) piece of symplectic potential in section 3.3. The charge contribution from this problematic piece will also appear to be divergent.  However, we will impose some restrictions on the behavior of free radiative data, $C_{AB}^{(0)}$, near $\mathcal{I}^+_{\pm}$ which will render it finite. Section 3.4 evaluates the charge contribution coming from non-problematic (finite as $t\rightarrow\infty$) portion of symplectic potential. In section 3.5, adding the charge contributions from the finite and divergent pieces of $\theta^t$, we will get the total soft charge. In the linearized Gravity under consideration, the hard charge only comes from the matter. It is evaluated and added to soft charge in section 3.6 to obtain the total asymptotic charge. It matches the charge expected from the leading soft graviton theorem in six dimensions. Section 3.7 discusses the generalized CK constraints in six dimensions. The calculation is done in the spirit of \cite{laddha}. 
 \subsection{Boundary Conditions and Constraints From Einstein's Equations} 
In six dimensions, the boundary conditions for asymptotically flat spacetimes are
\be \begin{split}
g_{uu}=-1+O(r^{-1}), \hspace{.2 in} g_{ur}=-1+O(r^{-2}), \hspace{.2 in} g_{uA}=O(1), \hspace{.2 in}
g_{AB}=r^2\gamma_{AB}+O(r), \\
R_{uu}=O(r^{-4}), \hspace{.4 in} R_{ur}=O(r^{-5}), \hspace{.4 in} R_{uA}=O(r^{-4}), \\R_{rr}=O(r^{-6}) \hspace{.4 in}
  R_{rA}=O(r^{-5}), \hspace{.3 in} R_{AB} = O(r^{-4}) \end{split}   
\ee 
Using linearized Einstein's equations (\ref{Ricci})
 in Bondi coordinates, we get the following constraints:
 
\be \label{M}
M^{(1)} =0 , \hspace{.5 in }M^{(2)}=-\frac12 D^AU_A^{(1)}.   
\ee
The $R_{rA}$ equations give 
\be \label{eeU}
U_A^{(0)}=-\frac16 D^BC_{BA}^{(-1)} , \hspace{.5 in} U_A^{(1)}=-\frac13 D^BC_{BA}^{(0)}, \hspace{.5 in} U_A^{(2)}=-\frac34 D^BC_{BA}^{(1)}.    
\ee
The $R_{AB}$ equations lead to
\be \label{eeC}
\partial_uC_{AB}^{(-1)}=0
\ee
$R_{uu}$ equations give
\be \label{eeT}
\frac12 [D^2- 2] M^{2}  + \partial_u D^a U_a^{(2)} + 2  \partial_u M^{(3)}  + T^{M(4)}_{uu} =0.   
\ee
where $T_{uu}^{M(4)}$ is the $\mathcal{O}(r^{-4})$ component of matter stress-energy tensor.
$C_{AB}^{(0)}$ is free, unconstrained data.
\subsection{Symplectic Potential} \label{sec3.1}

Let's now proceed with the derivation of asymptotic charge by employing the covariant phase space approach (see section \ref{covariant}). The first step is to compute the symplectic potential. 
 We will find that it is divergent as $t\rightarrow \infty$. \\ 
 We have (as discussed in \ref{covariant}),
\be
\theta^t= \theta^r+\theta^u.
\ee

Doing this computation  (whose details are deffered to appendix \ref{sp}), one gets 
\be 
 \theta^t=\frac12r^{4}\sqrt{\gamma}\bigg[\delta U^A g_{AB} \partial_r U^B -\delta g^{AB} \bigg(D_{(A}U_{B)}+\frac12 \partial_u g_{AB} \bigg)\bigg].
\ee

From the above equation, it is clear that $\theta^t$ has divergent terms since if one expands the metric components in powers of $r$, the terms inside the square brackets are of the order $r^{-3}$ while there is $r^{4}$ sitting outside the square brackets. Thus, there are divergences as $t\rightarrow \infty$ (since $r=t-u$). We are now going to compute the charge from $\theta^t$. We will see that there are no divergent contributions to the charge, even from the divergent piece of $\theta^t$, once suitable fall offs, in $u$, on the free radiative data ($C_{AB}^{(0)}$)  are imposed.\\

\emph{From now onwards, sphere indices will be raised and lowered with $\gamma_{AB}$}.
\subsection{Soft charge contribution from divergent part of $\theta^t$} \label{divergences}
It will be helpful to study the divergent pieces in the symplectic potential separately from the finite pieces and see how the right fall-offs near $\mathcal{I}^+_{\pm}$ kill the divergences and give some finite contribution to the charge.  \\

Expanding the metric components in powers of $r$ (section \ref{metric}), we find that the potentially divergent (with positive powers of $r$) part of $\theta^t$, $\theta^t_D$ (say), is
\begin{equation}
\theta^t_D= \frac{r^4}{2} \sqrt{\gamma}\bigg[\delta U_A^{(0)}  \partial_r\bigg(\frac{U^{A(0)}}{r^2}\bigg) +\frac{\delta C^{AB (-1)}}{r^3} \bigg(D_{(A}U_{B)}^{(0)} + \frac{r}{2}\partial_u   C_{AB}^{(-1)} + \frac12 \partial_u C_{AB}^{(0)} \bigg) \bigg]. 
\end{equation}

Now, using Einstein's equations (\ref{eeC}) and ignoring total variations, we have

\be
\theta^t_D=\frac{\sqrt{\gamma}}{2}\bigg[r\delta C^{AB(-1)}\bigg(D_{(A}U_{B)}^{(0)}+\frac12 \partial_u C_{AB}^{(0)}\bigg)\bigg].
\ee

Substituting for $U_B^{(0)}$ (\ref{eeU}), we see that the first term is a total variation. Total variations don't contribute to the symplectic form (or to the charge) since the symplectic form involves antisymmetrization of the variations. 
For the second term, substitute $r=t-u$. Then, upto total variations,
\be \label{eqthetad}
\theta^t_D=\frac14 \sqrt{\gamma} (t-u) \partial_u C^{(0)}_{AB}\delta C^{AB(-1)},
\ee


Integrating by parts in $u$,
\be
\theta^t_D=\frac14 \sqrt{\gamma}  C^{(0)}_{AB}\delta C^{AB(-1)}+\partial_u \bigg(\frac14 \sqrt{\gamma} (t-u)  C^{(0)}_{AB}\delta C^{AB(-1)}\bigg).
\ee


Let's call the contribution to the symplectic structure $\Omega^D$. Thus, for supertranslations, the form of variations (\ref{st}) lead to
\be
\Omega^D(\delta_{\xi}, \delta)=\lim_{t\rightarrow\infty} \bigg[-\int_{\mathcal{I^+}}\ \frac{\sqrt{\gamma}}{2} \delta C_{AB}^{(0)}D_A D_B f - \int_{\mathcal{I^+}}\ \partial_u \bigg((t-u)\frac{\sqrt{\gamma}}{2} \delta C_{AB}^{(0)}D_A D_B f\bigg)\bigg].
\ee

In the limit $t\rightarrow\infty$ the above expression is generically divergent. To avoid this, in the limit $t \to \infty$, we impose the condition
 \be  \label{bc} 
D^{A}D^{B}C^{(0)}_{AB}(u=+\infty, \, \hat{x})=D^{A}D^{B}C^{(0)}_{AB}(u=-\infty, \, \hat{x})= O(\vert u\vert^{-1-\epsilon})
\ee
where $\hat{x}$ labels a point on the conformal sphere. 

 Conditions (\ref{bc}) ensure the vanishing of the potentially divergent total derivative term after $u$-integration. 
 It should be noted that in the linearized regime, the fall-offs (\ref{bc}) are consistent with the action of supertranslations, i.e., if one starts with a portion of radiative phase space parametrized by the free radiative data ($C^{(0)}_{AB}$) of the form given by eq. (\ref{bc}) (along with non-radiative/"kinematical" data $C^{(-1)}_{AB}$), supertranslations preserve it. This can be easily seen in the linearized regime by recalling $\delta_{\xi}C_{AB}^{(0)}=0$ (\ref{st}). \footnote{\label{footnote}However, under the full action of supertranslations, i.e., including the pieces which are homogeneous in the fields and contribute to the hard part of the charge, the fall-offs are violated (more on this is discussed in section \ref{problems}). This is not satisfactory and one possible resolution may be to add  boundary counterterms 
  to the action to cancel the divergences instead of demanding the fall-offs (\ref{bc}).
  }

Imposing the above conditions (\ref{bc}) leaves us with a finite contribution to the supertranslations charge. We will denote it with superscript `D' to indicate that it is the contribution from the divergent (as $t\rightarrow\infty$) part of $\theta^t$.

\be \label{qd}
Q^D_{\xi}=-\int_{\mathcal{I^+}}\ \frac{\sqrt{\gamma}}{2} D_A D_BC_{AB}^{(0)} f.
\ee


\subsection{Soft charge contribution from the finite part of $\theta^t$}
The finite part of $\theta^t$ (terms with $r^0$) is
\begin{multline}
\theta^t_F=\frac{\sqrt{\gamma}}{2}\bigg[\delta U^{(1)}.U^{(0)}+\big(\delta C^{(-1)}.DU^{(1)}-\delta(DU^{(0)}).C^{(0)}\big) \\ +\frac12 \big(\partial_u C^{(1)}.\delta C^{(-1)}+\partial_u C^{(0)}.\delta C^{(0)}\big)\bigg]
\end{multline}
after ignoring total variations. Note that the last term doesn't contribute to the symplectic form for supertranslations since $\delta C_{\xi}^{(0)}=0$. Further simplifying, using (\ref{eeU}), terms contributing to charge are
\be
\theta_t^F=\sqrt{\gamma}\bigg[\delta C_D^{A(-1)}\bigg(-\frac1{18} D^DD.C_A^{(0)}+\frac14\partial_uC_{AB}^{(1)}\bigg)\bigg]
\ee

Using Einstein equations (in the absence of matter), 
$D.D.C^{(1)}=0$ (see \cite{campoleoni}), we get 

\be \label{qf}
Q_{\xi}^F=\int_{\mathcal{I^+}}\frac{\sqrt{\gamma} f}{3}\bigg(\frac{D^2}{4}+1\bigg) D^AD^BC^{(0)}_{AB} 
\ee
\subsection{Total soft charge}
From (\ref{qd}, \ref{qf}), the total soft charge for supertranslations is
\be
Q^{soft}_{\xi}=Q_{\xi}^D+Q_{\xi}^F= \int_{\mathcal{I^+}}\frac{\sqrt{\gamma} f}{12}\big(D^2-2\big) D^AD^BC^{(0)}_{AB}. 
\ee

\subsection{Total charge in linearized gravity}
Having found the soft charge, we now turn to hard charge. When there is coupling to matter, we assume the fall-offs of $T_{\mu\nu}$ to be same as those of $R_{\mu\nu}$ as in \cite{strominger}. Then, the hard contribution to the charge coming from matter is given by
\begin{equation}
 Q_{\xi}^{\text{hard}} =  \lim_{t\rightarrow\infty} \int_{\Sigma_{t}}r^{D-2}\mathrm{d}u\, \mathrm{d}\Omega_{D-2}\, n_{\Sigma_t}^{\mu}T^M_{\mu\nu}\xi^{\nu}.
\end{equation} 
which yields, for supertranslations, 
 \begin{equation}\label{hard-charge}
 Q_{\xi}^{\text{hard}} =   \int_{\mathcal{I}^+}\sqrt{\gamma} f(z) T^{M(4)}_{u u}.
\end{equation} 
Therefore, the total charge is 
\be
Q_{\xi}= Q_{\xi}^{soft}+Q_{\xi}^{hard}=\int_{\mathcal{I^+}}\frac{\sqrt{\gamma} f(z)}{12}\big(D^2-2\big) D^AD^BC^{(0)}_{AB} +\int_{\mathcal{I}^+} \sqrt{\gamma}f(z) T^{M(4)}_{u u}.
\ee
$Q_{\xi}$ matches with the soft charge obtained in \cite{strominger} by starting from the leading soft graviton theorem in six dimensions and writing it as a Ward identity.
\footnote{While comparing with \cite{strominger}, one should keep in mind that we are working in the units $8\pi G=1$ 
. Thus, for us, $\kappa^2=32\pi G= 4$.}

This can also be obtained as an `electric charge' 
using electric components of Weyl tensor as shown in the appendix \ref{electric}.

\subsection{Generalized CK constraints}
In four dimensions,  there are $d(d-3)/2=2$ leading soft graviton theorems, one corresponding to each helicity. But, we have only one canonical asymptotic charge corresponding to supertranslations; giving rise to only one soft theorem. The counting is rescued by the presence of one CK (Christoudoulou-Klainermann) constraint (\cite{CK}) which is necessary for a well-defined scattering problem \cite{strominger1}. 
Similarly, in six dimensions, there are $d(d-3)/2=9$ leading soft graviton theorems, 
. But we have only one canonical charge; giving rise to only one independent soft theorem. So, analogous to four dimensions, we expect that there should be eight generalized CK constraints in six dimensions. Six out of these were given in \cite{strominger} by the requirement of vanishing of the magnetic component of the Weyl tensor near boundaries of null infinity \footnote{In four dimensions, the origin of the CK constraints can be understood as the requirement that the magnetic charge (as defined in \cite{electricladdha}) vanish. However, in higher dimensions, their origin is not so well-understood in this way. A straightforward analog of this magnetic charge vanishes identically in higher dimensions giving no constraints (also see appendix \ref{electric}. }
\be
C_{urAB}(u=\pm \infty,\ \hat{x})=\bigO(r^{-2}).
\ee
The $\bigO(r^{-1})$ component of this equation gives 
\be \label{bcM}
D_{A}U_B^{(0)}-D_{B}U_A^{(0)}=0
\ee
It was pointed out in \cite{strominger} that this condition together with (\ref{eeC}) implies that $C^{(-1)}_{AB}$ is a pure supertranslation.
Now, we propose that `$u$' fall-offs (\ref{bc}) that were imposed to ensure finiteness of charge are a part of the CK constraints in six dimensions. This gives us two more CK type constraints. Together with (\ref{bcM}), they make up a total of 8 generalized CK constraints needed in six dimensions.

\section{Generalization to arbitrary even dimensions} \label{Charge}
In this section, we generalize our analysis to arbitrary higher even dimensions. We first give the conditions necessary for a finite, well-defined charge. After that, the strategy is the same as in six dimensions. We use covariant phase space analysis and first analyze the symplectic potential in arbitrary dimensions. It has divergences but we go ahead and compute the charge. The conditions mentioned in (\ref{conditions}) then ensure that it is finite and equal to the expected charge from the leading soft graviton theorem. 

\subsection{Conditions for finiteness of the charge} \label{conditions}
Einstein's equations determine $D.D.C^{(n)}$, $0<n<m-3$, recursively in terms of $D.D.C^{(-1)}$ and determine  $D.D.C^{(n)}$, $m-1<n<2m-3$, recursively in terms of $D.D.C^{(m-2)}$ 
(see \cite{campoleoni})
\be \label{eqC}
\partial_u D.D.C^{(n)}=\mathcal{D}_{n,m}D.D.C^{(n-1)},  \hspace{.99in} 0\leq n\leq m-3, \hspace{.2in} m-1\leq n\leq 2m-3
\ee
where
\be  \label{eqC1}
\mathcal{D}_{n,m}=\frac{n (2 m-n-3)}{2 (n+2) (-2 m+n+1) (-m+n+2)} \left(D^2-(n+1) (2 m-n-2)\right).
\ee\\
As is clear from the above equation $D.D.C^{(n)}$ is determined in terms of $D.D.C^{(n-1)}$ but upto an `integration constant', i.e.,  a function on the sphere. 
Now, we will, basically, demand the vanishing of these integration constants. Furthermore, we will impose two conditions on the fall-off behaviour of the free radiative data, $C^{(m-2)}$, near the (two) boundaries of $\mathcal{I}^{+}$. More concretely, the conditions are\\
\be \label{bcu1}
D.D.C^{(n)}= u^{n+1}\bigg( \prod_{j=0}^{n} \mathcal{D}_{j,m}\bigg)D.D.C^{(-1)},  \hspace{.99in} 0\leq n\leq m-3,
\ee\\
\be      \label{bcu2}  
D.D. C^{(m-2)}\bigg\vert_{(u=\pm\infty, \, \hat{x})}  \backsim \mathcal{O}(\vert u \vert^{-m+1-\epsilon)}), 
\ee\\
\be    \label{bcu3}    
D.D. C^{(m-2+n)}\bigg\vert_{(u=\pm\infty, \, \hat{x})}  \backsim \mathcal{O}(\vert u \vert^{-m+1+n-\epsilon)}), \hspace{.99in} 1\leq n \leq m-2,\; \epsilon>0.
\ee\\
These $\big((m-2)+2+(m-2)=2m-2\big)$ conditions, together, will ensure the finiteness of the charge as we shall see now.
\subsection{Symplectic Potential}

Calculating symplectic potential in arbitrary dimensions, throwing away unnecessary terms and using Einstein's equations, the part of $\theta^{t}$ that contributes to the soft charge is (for details see, appendix \ref{sp})

\begin{equation} \label{thetaa}
\theta^t=\frac{r^{2m}\sqrt{\gamma}}{2}\Sigma_{n=0}^{n=2m-3}\bigg[\frac{\delta C^{AB(-1)}D_AD.C^{(n)}_B}{r^{n+4}}opt_{n,m}+\frac12\frac{\delta C^{(-1)}}{r^{n+3}}.\partial_uC^{(n)}
\bigg],
\end{equation}
where 
\be \label{opt}
opt_{k,m}= \frac{(k+1) m}{(k+3) (2 m-1) (k-2 m+2)}\ .
\ee \\

For most values of $n$ (just by counting the power of $r$), generically, there are divergent terms in $\theta^t$ as $t\rightarrow\infty$. However, the conditions (\ref{bcu1}), (\ref{bcu2}), and (\ref{bcu3}) ensure that the charge is finite as we will now see.

\subsection{Soft Charge}
The soft charge can be computed as explained above (section \ref{covariant}). It will only have terms containing $D.D.C^{(n)}$ since it has to be a scalar, linear in $C^{(n)}_{AB}$, and because of tracelessness of $C^{(n)}_{AB}$ s. An easy computation, then gives
\begin{multline} \label{deltaQ}
\delta Q_{\xi}^D=\int_{\mathcal{I^+}}\frac{\sqrt{\gamma}}{2}\sum_{n=m-2}^{n=2m-3}\bigg[\frac{1-2m}{m}\frac{(D^2+2m)D.D.\delta C^{(n)}}{r^{-2m+n+4}}opt_{n,m}
-\frac{\partial_uD.D.\delta C^{(n)}}{r^{-2m+n+3}}
\bigg]f(z).
\end{multline}

Note that the sum starts from $m-2$. This is because of eq. (\ref{bcu1}), which gives $C^{(0<n<m-2)}_{AB}$ in terms of $C^{(-1)}_{AB}$, making the terms with $n<m-2$ (in \ref{thetaa}) total variations that don't contribute to the charge.\\

Now, expressing the charge in terms of the free radiative data, $C_{AB}^{(m-2)}$, using the Einstein's equations (\ref{eqC}) (see \ref{charge}), we have
\begin{multline}\label{divQ}
\delta Q^D_{\xi}=\int_{\mathcal{I^+}}\frac{\sqrt{\gamma}}{2}\sum_{n=m-2}^{n=2m-4}\bigg[\bigg(\frac{1-2m}{m}\frac{(D^2+2m)}{r^{-2m+n+4}}opt_{n,m}\\
\\
-\frac{2m-n-3}{r^{-2m+n+4}}\bigg)
\prod_{i=m-1}^{n}\bigg(\int du\,\mathcal{ {D}}_{i,m}\bigg)D.D.\delta C^{(m-2)}f(z)\bigg]
\end{multline}
where $\mathcal{ {D}}_{i,m}$ is a differential operator on sphere defined by eq. (\ref{eqC1}) and we take $\prod_{i=m-1}^{m-2}(\int du\,\mathcal{{D}}_{m-2,m}):=1$.
The charge appears to be divergent since $2m-4\geq n$ but let's use a generalization of the trick used in section \ref{divergences}. 
To this end, we substitute $r=t-u$ and integrate each term in (\ref{divQ}) by parts (multiple times) w.r.t $u$ untill all the factors of $r$ are removed. For instance,
\be \label{ibp}
\int du\frac{D.D.\delta C^{(m-2)}}{r^{-2m+n+4}}=(2m-n-4)!\int du\; I^{(2m-n-4)}(D.D.\delta C^{(m-2)})
\ee
where $I^{(2m-n-4)}=\bigg(\int du\bigg)^{(2m-n-4)}$ is an integral operator denoting the $(2m-n-4)$-th antiderivative of the argument with respect to $u$. All the boundary terms vanish by the conditions (\ref{bcu2}) and (\ref{bcu3}) which can alternatively be written as the following fall-off conditions near $\mathcal{I}^+_{\pm}$:

\be        \label{bcu}
I^{(k)}\big(D.D.\delta C^{(m-2)}\big)\bigg\vert_{(u=\pm\infty, \, \hat{x})}  \backsim \mathcal{O}(\vert u \vert^{-m+1+k-\epsilon)}), \hspace{.3in} 0\leq k \leq m-2,\; \epsilon>0.
\ee

Note that these boundary conditions are preserved by the supertranslations in the linearized regime in even $d>4$ (see footnote \ref{footnote} and section \ref{problems} ) because supertranslations don't act on $C_{AB}^{(n \geq 0)}$ (upto homogeneous term which lead to non-linear terms in the charge)(\ref{st}). \\

Now, from equations (\ref{divQ}) and (\ref{ibp}), after simplification (see \ref{charge}),  we obtain

\begin{multline} \label{Q}
Q_{\xi}^{soft} =  \frac1{(2m-1)} \frac{2^{-m}} {\Gamma(m)}    \int_{\mathcal{I}^+}   \sqrt{\gamma} \;f(z)  \prod\limits^{2m-1}_{l=m+1}  (D^2  -(2m-l)(l-1) )
 \; I^{(m-2)}(D. D.  C^{(m-2)}).
\end{multline}
The charge is same as in \cite{strominger} as can be seen using the integral property of Fourier transforms as follows.  Each antiderivative brings down a power of frequency 
\footnote{Fourier transform of an integral also  involves a boundary term which is basically the zero mode of the unintegrated function times a delta function. This term vanishes due to (\ref{bcu})}. The final $u$ integral over $\mathcal{I}^+$ picks out the zero mode of $(m-2)$-th antiderivative of $D.D.C^{(m-2)}$.

\subsection{Total Charge}
As in six dimensions, when we have matter coupled to the theory we get a hard matter contribution to the charge which is 
\be
Q_{\xi}=\int_{\mathcal{I}^+} \sqrt{\gamma}f(z) T^{M(2m)}_{u u}.
\ee
The total charge is, thus,
\begin{multline}
Q_{\xi}= Q_{\xi}^{soft}+Q_{\xi}^{hard}\\
= \frac1{(2m-1)} \frac{2^{-m}} {\Gamma(m)}    \int_{\mathcal{I}^+}   \sqrt{\gamma} \,f(z)  \prod\limits^{2m-1}_{l=m+1}  (D^2  -(2m-l)(l-1) )
 \; I^{(m-2)}(D. D.  C^{(m-2)})\\
 +\int_{\mathcal{I}^+} \sqrt{\gamma}f(z) T^{M(2m)}_{u u}
\end{multline}
This can also be obtained as an `electric charge' 
 using electric components of Weyl tensor as shown in the appendix \ref{electric}. 

\subsection{Generalized CK constraints}
In $d$ dimensions, there are $d(d-3)/2$ leading soft graviton theorems, one coresponding to each helicity.  But, we have only one canonical asymptotic charge coresponding to supertranslations; giving rise to only one independent soft theorem. So, we expect that there should be $d(d-3)/2-1$ generalized CK constraints  in $d$ dimensions. $(d-2)(d-3)/2$ out of these were given in \cite{strominger} by the requirement of vanishing of the magnetic component of Weyl tensor near boundaries of null infinity
\be
C_{urAB}(u=\pm \infty,\ \hat{x})=\bigO(r^{-2}).
\ee
The $\bigO(r^{-1})$ component of this equation gives 
\be \label{bcM1}
D_{A}U_B^{(0)}-D_{B}U_A^{(0)}=0
\ee
This condition togeher with (\ref{eeC}) implies that $C^{(-1)}_{AB}$ is a pure supertranslation.
Now, as in six dimensions we propose the connditions (\ref{bcu1}), (\ref{bcu2}), and (\ref{bcu3}) (which ensure finiteness of the charge) to be some of the CK constraints in arbitrary dimensions. This gives us $2m-2=d-4$ more CK type constraints. Together with (\ref{bcM1}), they make up the required total of \hspace{.1in} $d(d-3)/2-1\big(=(d-2)(d-3)/2+(d-4)\big)$ \hspace{.1in} generalized CK constraints needed in $d$ dimensions.

\section{Problems in going beyond linearized regime} \label{problems}
In this work, we have only restricted ourselves only to the linearized gravity coupled to scalar matter. There are several important issues in the full non-linear theory whose resolution is beyond the scope of this paper . To understand these issues, let us consider the case of six dimensions for concreteness. In the full theory, under supertranslations, the radiative data transforms as,
\begin{multline} \label{non-linear}
\delta C^{(0)}_{AB}=-\frac{C_{AB}^{(-1)}}{8} D^2 f+\gamma_{AB}\bigg(\frac23D^DfD.C_D^{(-1)}+\frac{D^AD^CfC^{(-1)}_{CD}}{2}\bigg)+\\\frac{D_AfD.C_B^{(-1)}}{6}-\frac{D_AD^CfC^{(-1)}_{CB}}{2}-f\frac{\partial_uC^{(0)}_{AB}}{2}.
\end{multline}
 
The right hand side of the above equation contains $u$ independent pieces (all the terms containing $C^{(-1)}$). Thus, this transformation violates the $u$ fall-off conditions that were imposed to ensure the finiteness of the soft charge (\ref{bc})
 \be   
D^{A}D^{B}C^{(0)}_{AB}(u=+\infty, \, \hat{x})=D^{A}D^{B}C^{(0)}_{AB}(u=-\infty, \, \hat{x})= O(\vert u\vert^{-1-\epsilon}).
\ee
Instead of imposing these fall-offs, it may be possible to settle this issue by adding some counterterms in the action, but this remains to be investigated. 
  
Moreover, it turns out that in full theory, generically there are quadratic terms in the charge which are functionals of $C^{(0)}_{AB}$ and $C^{(-1)}_{AB}$. Being linear in free radiative data $C^{(0)}_{AB}$, they would potentially contribute to the soft charge but as $C^{(-1)}$ parametrize vacuua of the theory, an understanding of such terms would require us to understand coupling of the ``kinematic data" $C^{(-1)}$ with radiative data $C^{(0)}$. In the absence of such understanding, our analysis of Asymptotic symmetries of full theory which is consistent with soft graviton theorem remains incomplete.


\section{Conclusions and outlook}

Supertranslation symmetries in higher even dimensions are in an ambivalent state. Detailed analysis in classical gravity literature suggests that Asymptotic symmetry group of general relativity in higher (even) dimensions is precisely the Poincare group. Primary reasons for this rather stringent constraint on the asymptotic structure of the theory are the following.  Unlike in four dimensions, supertranslations aren't tied to gravitational radiation (and more precisely the Memory effect)  in higher even dimensions.\footnote{The case of odd dimensions is far more subtle as in the presence of radiation, there does not exist a notion of null infinity and even formulating questions regarding Asymptotic symmetries seems far more difficult.\cite{arnabunpublished}} This enables one to choose boundary conditions that permit radiating solutions without allowing supertranslations.  Although in principle, one can relax these boundary conditions, such conditions were thought to lead to an ill-defined asymptotic charge due to divergent symplectic current. 

However, soft graviton theorems which in four dimensions are statements about (asymptotic) symmetries of the $\s$-matrix, are robust constraints on perturbative Quantum Gravity in higher dimensions. In fact, their status is far more transparent in higher dimensions due to the absence of Infra-red divergence and well-definedness of the perturbative $\s$-matrix.  In \cite{strominger}, a strong evidence was given for the existence of supertranslations in higher even dimensions precisely by writing the leading soft graviton theorem as a Ward identity of \emph{the} supertranslation symmetry. What has however been absent so far in the literature is an analysis of relaxed boundary conditions in classical Gravity which admits Supertranslations \emph{and} lead to a well-defined asymptotic charge at ${\cal I}$, whose Ward identities are precisely the one formulated in \cite{strominger}. We have taken certain steps towards filling this gap in the paper. 

Namely, we obtain the (asymptotic) charge for supertranslations in arbitrary even dimensions using covariant phase space approach in the \emph{linearized} gravity coupled to massless matter. We have shown that this charge is well-defined and matches the results expected from the leading soft graviton theorem \cite{strominger}. One may wonder, how does our analysis (albeit in the context of linearized perturbations) bypass the rigorous no-go results obtained in the literature. That is,  the use of relaxed boundary conditions proposed in \cite{strominger} allows for supertranslations. This, however, appears to lead to an ill-defined charge in accordance with the analysis in \cite{Wald}. We cured this by \\
\noindent $\bullet$  Imposing appropriate fall-offs in `u' for the free radiative data near the boundaries of $\mathcal{I^+}$ ($\mathcal{I}^+_{\pm}$) (\ref{bcu2}).\\
$\bullet$ Imposing conditions for the vanishing of `integration constants' (\ref{bcu1}, \ref{bcu3}) that appear while solving Einstein's equations (\ref{eqC}).

%
We further proposed these conditions to be the generalized CK-constraints along with the conditions of vanishing of the magnetic part of the Weyl tensor (\ref{bcM1}) near $\mathcal{I}^+_{\pm}$. 
They were shown to match the number of constraints required to establish the uniqueness of leading soft graviton theorem. 
One may worry about whether or not the `$u$' fall-offs (\ref{bcu2}) allow for interesting physical situations.  
 With regards to this, it can be easily checked that at least the gravity waves radiated from the classical scattering processes (which were studied in \cite{sen}) respect these fall offs. We expect them to be satisfied in other physically relevant scenarios as well.  \\
 
 However, there are some important open issues that remain open and that pertains to the analysis of Asymptotic symmetries in full (as opposed to linearized) theory. Once higher order perturbations are allowed, supertranslations do not appear to respect the $u$- fall-offs imposed on the radiative data (\ref{non-linear}). This could possibly be settled by adding boundary counter-terms to the action instead of imposing the $u$ fall-offs. However, this issue requires further investigation. 

Unlike in four dimensions where the origin of so-called CK constraints is well understood and is tied to vanishing of the asymptotic charge associated to the magnetic part of Weyl tensor, no such interpretation seems available in higher dimensions. It will also be interesting to see if our analysis can be generalized to include massive fields along the lines of \cite{laddhamassivegravity}.

It may also be noted that in six dimensions, the structure of the potentially problematic (divergent) terms and the  $u$ fall-offs imposed to cure them is similar those in  \cite{laddha} in the case of subleading soft graviton  theorem  in  four  dimensions.  This  hints  at  some  possible  connection between  subleading  soft  graviton  theorems  in  lower  dimensions  and  leading  soft  graviton  theorems  in  higher  dimensions  as  was  mentioned  in  \cite{stromingermemory}.
 The case of superrotations in higher even dimensions is under investigation.\\
 
 
 \section*{Acknowledgements}
I am very grateful to Alok Laddha for suggesting this problem;  his guidance, insightful discussions and numerous contributions throughout, including his critique of the final draft,  made this project possible. I would also like to thank Arnab Priya Saha for clarifying many basic and subtle points during several illuminating discussions that we had at the early stages of the project, suggesting the very crucial reference \cite{campoleoni}, and sharing his ongoing work in odd dimensions \cite{arnabunpublished}.
\newpage 
 \appendix
 \section{Details of charge computation in higher dimensions}
\subsection{Symplectic Potential} \label{sp}
Details of this section are valid in all even dimensions. We have
\be
\theta^r :=\frac{1}{2} \sqrt{g} \big( g^{bc} \delta \Gamma^r_{bc} - g^{r b} \delta \Gamma^c_{c b} \big).
\ee
Using linearised version of Christoffel symbols in \cite{barnich} \footnote{Although, in \cite{barnich}, they were written in the context of four dimensions, they are valid in all dimensions.} (with $\beta=0$ in linearized theory) we get

\begin{multline} 
\theta^r=\frac12r^{2m}\sqrt{\gamma}\bigg[\delta\big(\partial_r M\big)-U^A\delta\big(g_{AB} \partial_r U^B\big)+ g^{AB} \delta\bigg(D_{(A}U_{B)}+\frac12 \partial_u g_{AB} + \frac{M}{2} \partial_r g_{A B}\bigg)\bigg] 
\end{multline}
Neglecting terms which are total variations (since they do not contribute to the symplectic structure), we have

\be 
\theta^r=\frac12r^{2m}\sqrt{\gamma}\bigg[\delta U^A g_{AB} \partial_r U^B -\delta g^{AB} \bigg(D_{(A}U_{B)}+\frac12 \partial_u g_{AB} + \frac{M}{2} \partial_r g_{A B}\bigg)\bigg] 
\ee
Similarly,
\be
\theta^u :=\frac{1}{2} \sqrt{g} \big( g^{bc} \delta \Gamma^u_{bc} - g^{u b} \delta \Gamma^c_{c b} \big),
\ee

\be 
\theta^u=\frac14r^{2m}\sqrt{\gamma}\delta\big(\partial_r g_{A B}\big). 
\ee

\begin{multline} 
\implies \theta^t=\frac12r^{2m}\sqrt{\gamma}\bigg[\delta U^A g_{AB} \partial_r U^B -\delta g^{AB} \bigg(D_{(A}U_{B)}+\frac12 \partial_u g_{AB} + r(M+1) \gamma_{A B}\bigg)\bigg].
\end{multline}
Here, we have kept only terms that are quadratic in metric perturbations. This is because we are working in the linearized regime 
.  Tracelessness of $\delta g_{AB}$ now implies
\be 
 \theta^t=\frac12r^{2m}\sqrt{\gamma}\bigg[\delta U^A g_{AB} \partial_r U^B -\delta g^{AB} \bigg(D_{(A}U_{B)}+\frac12 \partial_u g_{AB} \bigg)\bigg].
\ee

Now, we use the fact that $\delta_{\xi}C^{(n\geq0)}_{AB}=0$ to throw away terms which are of the form $\delta C^{(n\geq 0)} \delta C^{(m\geq 0)}$ since they don't contribute to the charge along with the total variations. We thus get the part of $\theta^t$ that contributes 
\begin{multline}
\theta^t=\frac{r^{2m}\sqrt{\gamma}}{2}\Sigma_{n=0}^{n=2m-3}\bigg[\frac{-(n+1)}{r^{n+4}} \delta U^{(0)}.U^{(n+1)}+\frac{\delta C^{(-1)}}{r^{n+4}}.DU^{(n+1)}\\
+\frac{\delta C^{(n)}}{r^{n+4}}.DU^{(0)}+\frac12\frac{\delta C^{(-1)}}{r^{n+3}}.\partial_u C^{(n)}
\bigg]
\end{multline}

Now, Einstein's equation for $R_{rA}$ gives
\be
U_A^{(n)} = f_{n,m} D.C_{A}^{(n-1)}, \hspace{.5 in} 0\leq n \leq 2m-2.   
\ee
where
\be
f_{n,m}=\frac{(n+1)}{(n+2)(n+1-2m)}.
\ee

Substituting in $\theta^t$, we have eq (\ref{thetaa})
\begin{equation}
\theta^t=\frac{r^{2m}\sqrt{\gamma}}{2}\Sigma_{n=0}^{n=2m-3}\bigg[\frac{\delta C^{AB(-1)}D_AD.C^{(n)}_B}{r^{n+4}}opt_{n,m}+\frac12\frac{\delta C^{(-1)}}{r^{n+3}}.\partial_uC^{(n)}
\bigg],
\end{equation}
where 
\be
opt_{k,m}= (k + 1) f_{k + 1, m} f_{0, m} + f_{1 + k, m} - f_{0, m}=\frac{(k+1) m}{(k+3) (2 m-1) (k-2 m+2)}\ .
\ee

\subsection{Soft Charge} \label{charge}
Starting from (\ref{deltaQ}), 
using the Einstein's equations, the rest  $D.D.C^{(2m-3> n\geq m-2)}$ can be expressed in terms of $D.D.C^{(m-2)}$ (see \cite{campoleoni}). In the absence of matter, $D.D.C^{(2m-3)}=0$ . Other relations are as given by 

\be 
\partial_u D.D.C^{(n)}=\mathcal{D}_{n,m}D.D.C^{(n-1)},  \hspace{.99in} 0\leq n\leq 2m-3
\ee
where
\be 
\mathcal{D}_{n,m}=\frac{n (2 m-n-3)}{2 (n+2) (-2 m+n+1) (-m+n+2)} \left(D^2-(n+1) (2 m-n-2)\right).
\ee

\be
\implies D.D.C^{(n)}=\int du\, \mathcal{D}_{n,m}D.D.C^{(n-1)}.  
\ee

\be
\implies D.D.C^{(n)}=\prod_{i=m-1}^{n}\bigg(\int  du\,\mathcal{{D}}_{i,m}\bigg)D.D.C^{(m-2)}, \hspace{.3in} n\geq m-1
\ee
 
A few things are clear here. 
The last term in (\ref{deltaQ}) can be integrated by parts in $u$ and the boundary term,  thus obtained, will vanish for the conditions (\ref{bcu2}) and (\ref{bcu3}). With that and using the Einstein's equations above, we have (\ref{divQ}).\\

Now, from  (\ref{divQ}) and (\ref{ibp}) we obtain
\begin{multline} \label{finQ}
Q_{\xi}^{soft}=(2m-n-4)!\int_{\mathcal{I^+}}\frac{\sqrt{\gamma}}{2}\sum_{n=m-2}^{2m-4}\bigg[\bigg(\frac{1-2m}{m}(D^2+2m)opt_{n,m}\\
\\
-(2m-n-3)
\bigg)\prod_{i=m-1}^{n}(\mathcal{D}_{i,m})I^{(m-2)}(D.D.C^{(m-2)})f(z)\bigg]
\end{multline}

where $opt_{n,m}$ and $\mathcal{D}_{i,m}$ are given by (\ref{opt}) and (\ref{eqC1}) respectively. This tedious looking expression factorises and simplifies \footnote{One should be able to prove the equivalence of (\ref{Q}) with (\ref{finQ}) by induction. Although we haven't been able to do so, verification  for any particular $m$ is easily done using a symbolic manipulation software like Mathematica. We will obtain (\ref{Q}) in another way by computing Electric charge as in case of 6D.} to (\ref{Q})
\begin{multline} 
Q_{\xi}^{soft} =  \frac1{(2m-1)} \frac{2^{-m}} {\Gamma(m)}    \int_{\mathcal{I}^+}   \sqrt{\gamma} \;f(z)  \prod\limits^{2m-1}_{l=m+1}  (D^2  -(2m-l)(l-1) )
 \; I^{(m-2)}(D. D.  C^{(m-2)}).
\end{multline}

\section{Electric Charge} \label{electric}
Analogus to \cite{laddha}, we define electric charges \footnote{\cite{electricladdha} also defines a magnetic charge which gives the CK constraint in four dimensions. Its generalisation in higher even dimensions vanishes identically (atleast in the linearized regime).} 
\begin{eqnarray}
Q_{\mathcal{I}}[\xi] & :=& \lim_{t \to \infty} \int_{\Sigma_t} \partial_a (\mathcal{E}^a_{\ b}\xi^b)  \label{QEgen} 
\end{eqnarray}
where
\be
\mathcal{E}^{a}_{b}\ := -\frac{1}{2m-1}r \, \sqrt{g} \, C^{at}_{\phantom{at} br}.
\ee 

Here the relevant components of the Weyl tensor are
\be
C^u_{\phantom{u}uur}=-\frac12\partial_r^2M
\ee
and
\be
C^u_{\phantom{u}uAr}=\frac12\bigg(\partial_r-\frac1{r}\bigg)\partial_rU_A.
\ee
Then the electric charge defined by (\ref{QEgen}) gives
\be
Q_{\mathcal{I}}[\xi]= m\int_{\mathcal{I}^+}\partial_uM^{(2m-1)}f(z)
\ee
where we have used Einstein's equations and (\ref{bcu}). Writing it out in terms of the free radiative data using Einstein's equations, we see that, the charge obtained is the same as the one above which was obtained using symplectic phase space formalism
\begin{multline} 
Q_{\mathcal{I}}[\xi]= Q_{\xi}= \frac1{(2m-1)} \frac{2^{-m}} {\Gamma(m)}    \int_{\mathcal{I}^+}   \sqrt{\gamma} \,f(z)  \prod\limits^{2m-1}_{l=m+1}  (D^2  -(2m-l)(l-1) )
 \; I^{(m-2)}(D. D.  C^{(m-2)})\\ 
 +\int_{\mathcal{I}^+} \sqrt{\gamma}f(z) T^{M(2m)}_{u u}.
\end{multline}



\section{Commutators} \label{com}
Some useful commutation relations used above are

\begin{eqnarray}
 \left[D^{B}, D_{N}\right]C_{AB}^{(0)} &=& 2mC_{AN}^{(0)} \nonumber\\
 \left[D^{B}, D_{M}\right]D_{N}C_{AB}^{(0)} & = & D_{M}C_{AN}^{(0)} + 2mD_{N}C_{AM}^{(0)} - q_{MN}D.C_{A}^{(0)} \nonumber\\
 \left[D^{A}, D_{N}\right]D.C_{A}^{(0)} & = & (2m-1)D.C_{N}^{(0)} \nonumber\\
 \left[D^{A}, D_{M}\right]D_{N}D.C_{A}^{(0)} & = & D_{M}D.C_{N}^{(0)} - \gamma_{MN}D.D.C^{(0)} + (2m-1)D_{N}D.C_{M}^{(0)}.
\end{eqnarray}

\newpage 
\bibliographystyle{utphys}
\bibliography{STreferences}

\end{document}